\documentclass[aps,prb,twocolumn]{revtex4}
\usepackage{graphicx}
\usepackage{amsmath}

\newcommand{\nn}{\nonumber}
\begin{document}

\title[]{Where will a pen fall to?} 


\author{Veniamin A. Abalmassov$^{1, 2}$}
\author{Dmitri A. Maljutin$^{1, 3}$}%
\affiliation{$^{1}$ Novosibirsk State University, Pirogova 2,
630090 Novosibirsk, Russia}%
\affiliation{$^{2}$ Institute of Semiconductor Physics SB RAS, Lavrentieva
13, 630090 Novosibirsk, Russia}%
\affiliation{$^{3}$ Budker Institute of Nuclear Physics SB RAS,
Lavrentieva 11, 630090 Novosibirsk, Russia}


\date{\today}%

\begin{abstract}

We propose a simple experiment that everybody can carry out just while
reading this paper. The only thing we need is a pen or a pencil and a
finger. The dynamics of the falling pen, which is in touch with the
finger, depends essentially on the initial inclination angle of the long
axis of the pen. We point out two different types of trajectories of the
falling pen and we investigate initial conditions for each of them to be
realized. We write differential equations describing the dynamics of the
pen and analyze them both numerically and analytically.
\end{abstract}

\newenvironment{dedication}
  { \begin{flushleft} \em \small}
  {\end{flushleft} }
\begin{dedication}
Dedicated to Serova Nelli Pavlovna, \\School Teacher in Physics
\end{dedication}

\maketitle

\section{Setup and experiment}

\begin{figure}[bl]
\centering
\includegraphics[width=0.30\textwidth]{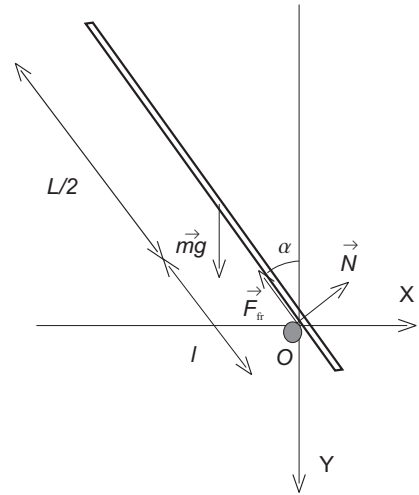}
\caption{The pen and a finger are in contact at the point {\it O}. We show
also a coordinate system, forces, and basic parameters which are defined
in the text.} \label{fig:init}
\end{figure}

We would like to show in our paper how sometimes very usual things taken
from our daily live can be formulated into an elegant physical problem.
Let us carry out a simple experiment. We take an ordinary pen by one of
its ends while keeping the other in touch with our finger as it is shown
schematically in Fig. \ref{fig:init}. At first, we allow for a large
enough initial angle between the long axis of the pen and the vertical,
say $\alpha_0 = \pi/4$. We let the pen fall down. Due to the contact with
the finger, however, it is not a free fall. While falling down the pen
slides across the finger and this makes it to rotate. Finally, it fell on
one side from the finger (for large initial angles, it is a side where the
center of mass of the pen was located initially). If then we decrease
gradually the initial angle and repeat the experiment again, at a certain
angle we notice that the center of mass of the pen starts to pass through
the point of the "pen-finger" contact and the pen falls down on the other
side from the finger.

The question is: What is the critical value of the initial
inclination angle in the considered setup which does separate the
two different types of motion of the pen in our experiment?

First, let us understand upon what this critical angle, $\alpha
_{\text{cr}}$, could depend? After some trials we can see that it depends
on the initial position of the contact point between the pen and the
finger. More precisely, being dimensionless, the critical angle depends on
the ratio of the initial distance between the center of mass of the pen
and the contact point to its total length, $l_0/L$. Interestingly, this
ratio together with the coefficient of kinetic friction, $\mu$, between
the pen and the finger, which is also dimensionless, are the only
parameters which determine the critical angle (we will consider a pen and
a finger of zero thickness which is close to realistic conditions and
simplifies essentially the problem). Indeed, it is not difficult to see
that we cannot construct any other dimensionless parameter in our problem,
for example, from the mass of the pen or the gravitational constant, which
means that the critical angle does not depend upon them.

We would like to determine the critical angle for all values of the
parameter $l_{0}$. However, this proves to be rather a hard task. We will
solve it only numerically. Two limits are of particular interest in this
problem that correspond to the case of $l_{0}=0$ and $l_{0}=L/2$. It is
interesting to test our intuition and to try to guess the value of the
critical angle in these limits. Is it $\pi/2$ and $0$ respectively or not?
We will try to exploit both of these cases analytically.

\section{Equations of motion}
\label{equations}

Differential equations which describe the evolution of the solid body
momentum, ${\bf P}$, and angular momentum, ${\bf M}$, can be written in
the general form \cite{LL1}:
\begin{equation}\label{equation-motion}
    \frac{d{\bf P}}{dt} = {\bf F}, \qquad
    \frac{d{\bf M}}{dt} = {\bf K},
\end{equation}
where ${\bf F}$ is the total external force applied to the body and ${\bf
K}$ is the total torque.

There are two independent variables in our problem: the angle $\alpha$ and
the length $l$, see Fig. \ref{fig:init}. However, the normal force, ${\bf
N}$, which acts on the pen at the pen-finger contact point and which is
perpendicular to the pen long axis will also enter our equations of
motion. So we will need three equations in total to describe the dynamics
of the pen. It is convenient to derive two of these equations from the
angular momentum evolution equation written with respect to two different
axes. First, we write this equation with respect to the axis passing
through the center of mass of the pen perpendicular to the plane of Fig.
\ref{fig:init} (in order to exclude the weight force from consideration):
\begin{equation}\label{ang-mom-center}
    \frac{d}{dt}\left[I_{0} \omega(t)\right]= N(t)l(t),
\end{equation}
where $I_{0}$ is the proper moment of inertia of the pen, $\omega(t)$ its
angular velocity, $N(t)$ the absolute value of the normal force, $l(t)$
the lever of this force. The dependence on time for quantities we indicate
by the parentheses on the right with the time variable, $t$, inside.

Now, we consider the evolution of the angular momentum defined with
respect to the axis passing through the contact point $O$ perpendicular to
the plane of Fig. \ref{fig:init} (we exclude the normal force from the
equation in this case):
\begin{equation}\label{ang-mom-contact}
    \frac{d}{dt}\left[(m l^{2} (t) + I_{0} )\omega(t)\right]
    = m g \, l(t) \sin \alpha(t),
\end{equation}
where $m$ is the mass of the pen, $g$ the gravitational constant.

The last equation we will obtain from the evolution equation for the total
momentum of the pen. For this purpose we choose the frame of reference
which rotates together with the pen, contact point $O$ being in its
origin. In this frame of reference the center of mass moves along one of
two coordinate axes which coincides with the pen's long axis. Since this
frame of reference is not inertial we should take into account the
centrifugal force acting on the pen as well. Finally, we have:
\begin{equation}\label{momentum}
    \frac{d}{dt}[m v_l (t)] = m l(t)\,\omega^2 (t)
    - m g \cos \alpha(t) + F_{\text{fr}}(t),
\end{equation}
where $v_{l}(t)$ is the velocity of the center of mass along the pen's
long axis, $F_{\text{fr}}(t)=\mu N(t)$ the friction force.

We will now simplify the above equations. First, we transform Eq.
(\ref{ang-mom-contact}) to obtain:
\begin{equation}\label{omega-deriv}
    \frac{d \omega(t)}{dt} = \frac{m l(t)}{m l^2 (t) + I_0}
    \left[g \sin \alpha(t) - 2 \,\frac{d l(t)}{dt}\,\omega(t)\right].
\end{equation}
Then, we find from Eq. (\ref{ang-mom-center}) the normal force:
\begin{equation}\label{normal-force}
    N(t) = \frac{I_0}{l(t)} \frac{d\omega(t)}{dt}
\end{equation}
and insert it in  Eq. (\ref{momentum}) which takes now the form:
\begin{align}\label{long-velocity}
    \frac{dv_l (t)}{dt} &= l(t) \omega^2 (t) - g \cos \alpha(t)
    \nn \\
    &+ \frac{\mu I_0}{m l^2 (t) + I_0}
    \left[g \sin \alpha(t) - 2 \,\frac{d l(t)}{dt}\,\omega(t)\right].
\end{align}

We note that by definition $\omega(t)=d\alpha/dt$ and $v_{l}(t)=dl/dt$.
Thus, we rewrite Eqs. (\ref{omega-deriv}) and (\ref{long-velocity}) and we
obtain the system of two coupled nonlinear second order differential
equations which describes the evolution of the angle and the center of
mass position in time:
\begin{align}
    \alpha '' &= \frac{l}{l^2 + I_0} \left(\sin \alpha - 2l'\alpha '
    \right), \label{alpha2} \\
    l'' &= l \alpha '^2 - \cos \alpha + \frac{\mu I_0}{l^2 + I_0}
    \left(\sin \alpha - 2l' \alpha '\right). \label{l2}
\end{align}
Here we have made all quantities dimensionless upon the substitutions: $t
:= t \sqrt{g/L}$, $l := l/L$, and $I_0 := I_0 /(mL^2)$. We use a prime
instead of $d/dt$ from now on in our formulas in order to make them more
compact.

We require that the solution of this system would satisfy four initial
conditions: $\alpha(0) = \alpha_0 $, $l(0) = l_0$, $\alpha '(0) = 0$, and
$l'(0) = 0$. The last two conditions correspond to the zero initial
velocity of the pen.

To find a solution of the above system is not a simple task. One way
consists in employing numerical methods. In the next Section we will show
results which gives this approach. After that we will consider two cases
when the system can be simplified and investigated analytically.

In Appendix {\ref{alt-way}} we give an alternative way of derivation of
the above system of equations specifically for those vigilant readers who
could feel themselves uncertain about the rotating frame of reference.

\section{Numerical simulation}

In order to solve numerically the system of Eqs. (\ref{alpha2}) and
(\ref{l2}) we first come back to variables $\omega$ and $v_{l}$ and we
rewrite it as a system of four first order differential equations with
four independent variables: $\alpha$, $\omega$, $l$ and $v:=v_{l}$. Then,
we use the finite difference method, namely Euler's method, to solve it.
For this we divide the time interval $[0, T]$ into $N$ small intervals,
each of a length $\Delta t = T/N$, and substitute, e.g., the derivative $d
\alpha (t)/dt$ by the ratio $[\alpha(t_{i+1})-\alpha(t_{i})]/\Delta t$,
where $t_{i+1}=t_{i}+\Delta t$ and $\alpha(t_{0})=\alpha_{0}$. So,
starting from $i=0$ we have at each step:
\begin{align}\label{num-syst}
    \omega_{i+1} &= \omega_{i} + \Delta t\,
    \frac{l_i}{l_i^2 + I_0} \left(\sin \alpha_i - 2v_i\omega_i
    \right),  \nn\\
    v_{i+1} &=v_{i} + \Delta t\!
    \left[l \omega_i^2 - \cos \alpha_i + \frac{\mu I_0}{l_i^2 + I_0}
    \left(\sin \alpha_i - 2v_i \omega_i \right)\right]\!\!,\nn\\
    \alpha_{i+1} &= \alpha_{i} + \Delta t\, \omega_{i}, \nn\\
    l_{i+1} &= l_{i} + \Delta t\, v_{i}.
\end{align}

\begin{figure}[t]
\centering
\includegraphics[width=0.3\textwidth]{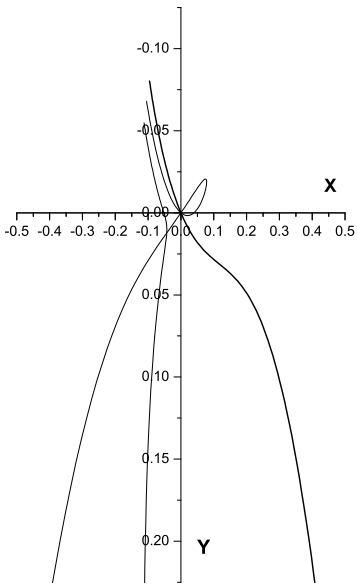}
\caption{Trajectories of the center of mass of the pen for three different
initial positions.} \label{fig:traject}
\end{figure}


Surprisingly, this simple numerical scheme turns out to be stable and
works very well. We show in Fig. \ref{fig:traject} the results of our
calculation for the trajectories of the pen's center of mass for three
different initial positions. We use the value of the proper moment of
inertia for a homogeneous stick $I_{0}=1/12$. We note a complex behavior
of the pen when its center of mass passes through the contact point twice.
In two other cases the pen falls explicitly on different sides from the
finger.

In order to determine the critical angle we simulate the fall of the pen
for a given initial length $l_{0}$. First, we take the initial angle
$\alpha_{0}$, to be equal to $\pi /4$ and calculate the trajectory of the
pen. If the pen fell on the left side from the finger (we just look at the
position of its center of mass at a height $y=0.1$) we subtract from
$\alpha_{0}$ its half and repeat the calculation. Otherwise, if the pen
fell on the opposite side, we add to the critical angle the half of it. We
repeat the cycle about ten times increasing or decreasing the initial
angle at each step by a half of an increment from a previous step. This
procedure allows to achieve a very good precision of the critical angle
value of about $1/2^{10} \sim 10^{-3}$. The above method lies in the class
of dichotomy methods.

\begin{figure}[b]
\centering
\includegraphics[width=0.35\textwidth]{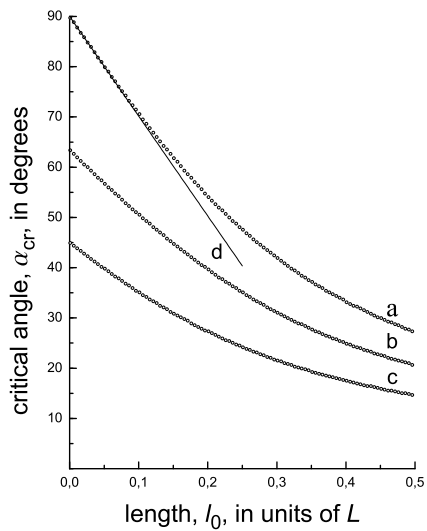}
\caption{Numerical results for the critical angle for three different
coefficients of friction: a) $\mu =0$, b) $\mu =0.5$, and c) $\mu =1$.
Line d) corresponds to the analytical solution in the case of critical
angles close to $\pi/2$.} \label{fig:result}
\end{figure}

We show in Fig. \ref{fig:result} the calculated critical angle values for
the whole range of the initial length $l_{0}$. We have carried out the
calculation for three values of the coefficient of kinetic friction $\mu$.
In the case of non zero value of the coefficient we see that
$\alpha_{\text{cr}}$ is smaller than $\pi /2$ when $l_{0}=0$. This is due
to the fact that there is no sliding for the angles $\alpha > \pi/2 -
\arctan \mu$, when the friction is available, which follows from the
couple of equations at equilibrium: $m g \cos\alpha =\mu N$ and $m g
\sin\alpha = N$. Thus, when $\alpha > \pi/2 - \arctan\mu$ the pen first
rotates without sliding, its center of mass falls below the finger and
then (when $\alpha > \pi/2 + \arctan\mu$) it starts to slide. In contrast,
when $\alpha < \pi/2 - \arctan\mu$ the pen starts to slide from the very
beginning. The numerical result for $l_{0}=0$, see Fig. \ref{fig:result},
agree well with the critical angle value $\alpha_{\text{cr}} = \pi/2 -
\arctan\mu$.


A peculiar feature of the dependence of $\alpha_{\text{cr}}$ on $l_{0}$ is
its finite value for $l_{0}=1/2$, i.e. when initially the pen touches the
finger by its very end. This angle is about $27^{\circ}$ for $\mu =0$.

\section{Analytical solution}

We consider now two cases when the system of equations (\ref{alpha2}) and
(\ref{l2}), which describes the evolution of the inclination angle and the
center of mass position of the pen in time, can be substantially
simplified.

\subsection{Angles close to $\pi/2$}

It should be clear that the critical angle will be about $\pi/2$ when
$l_{0}$ is close to zero. In the limit $\vert l\vert \ll \sqrt{I_0}$ and
$\vert \gamma \vert \ll 1$, where $\gamma = \pi/2 - \alpha$, Eqs.
(\ref{alpha2}) and (\ref{l2}) simplify to the system (we consider only the
case $\mu =0$):
\begin{align}
    I_0 \gamma '' &= - l, \label{gam2deriv}\\
    l'' &= - \gamma. \label{l2deriv}
\end{align}
We have supposed the derivatives $\gamma '$ and $l'$ to be negligibly
small compared to unity as well. The solution of this system is a
combination of trigonometric and hyperbolic sines and cosines. However,
due to the zero initial condition for the velocity the sines drop out of
the solution. Thus, we have
\begin{align}\label{solution-large-angl}
    \gamma (t) &= A \cos \omega t + B \cosh \omega t, \\
    l(t) &= A \sqrt{I_{0}} \cos \omega t - B \sqrt{I_{0}} \cosh \omega t,
\end{align}
where $A$ and $B$ are constants and $\omega = I_{0}^{-1/4}$.

We suppose that the critical trajectory corresponds to the case when $l(T)
= 0$ and $\gamma(T) = 0$ for a particular moment $T$. This could probably
contradict the observation that the center of mass can pass twice through
the contact point (see Fig. \ref{fig:traject}). However, we can hope that
our assumption is valid at least for the limit $\gamma \ll 1$. It results
in $B=0$ and the oscillatory solution for $\gamma$ and $l$:
\begin{align}\label{solution-large-angl}
    \gamma (t) = \gamma_{0} \cos \omega t \quad \text{and} \quad
    l(t) = l_{0} \cos \omega t,
\end{align}
where $\gamma_{0}=l_{0}/\sqrt{I_{0}}$, which is exactly the condition we
searched for. Finally, the critical angle in the considered limit equals
to $\alpha_{\text{cr}}=\pi/2 - l_{0}/\sqrt{I_{0}}$. We trace this
dependence of $\alpha_{\text{cr}}$ upon $l_{0}$ in Fig. \ref{fig:result}.
The agreement between our theoretical result and numerical data for angles
$\alpha_{\text{cr}}$ close to $\pi/2$ is quite remarkable.


\subsection{Small angles}

Another limit of interest corresponds to small angles $\alpha$. The
principal question is whether it is possible to prove analytically that
for $l_{0}=1/2$ the critical angle $\alpha_{\text{cr}}$ has a finite
non-zero value. So, we consider small angles $\alpha \ll 1$ and $l_{0}$
close to $1/2$ and we rewrite Eqs. (\ref{alpha2}) and (\ref{l2}) in this
limit as:
\begin{align}
    &(l^2 + I_0)\alpha '' + 2 l l'\alpha '
    - l \alpha = 0, \label{alpha2-small-angl} \\
    &l'' = -1. \label{l2-small-angl}
\end{align}
The solution of Eq. (\ref{l2-small-angl}) is $l = -t^2/2 + l_{0}$ which
corresponds to a free fall of the center of mass. In Eq.
(\ref{alpha2-small-angl}) we switch from the time variable, $t$, to the
length variable, $l$, arriving at the equation:
\begin{align}\label{alpha-l-small-angl}
    \frac{d^2 \alpha}{d l^2}
    + \left[ \frac{2 l}{l^2 + I_{0}} - \frac{1}{2(l-l_{0})}\right]
    \frac{d \alpha}{d l}
    -\frac{l\alpha}{2(l^2+I_{0})(l_{0}-l)} =0.
\end{align}

We are not able to fully solve this equation. However, if we consider only
the very beginning of the movement, when $(l_{0}-l) \ll 1$, we need to
keep only third and fourth terms in Eq. (\ref{alpha-l-small-angl}) which
simplifies it essentially:
\begin{align}\label{alpha-l0-small-angl}
    \frac{d \alpha}{d l}
    +\frac{l\alpha}{l^2+I_{0}} =0.
\end{align}
The solution of the above equation is
\begin{align}\label{alpha-small-angl-sol}
    \alpha (l)=\alpha_{0}\sqrt{\frac{l_{0}^2+I_{0}}{l^2+I_{0}}},
\end{align}
which fulfills initial conditions $\alpha=\alpha_{0}$ and $\alpha'=0$ at
$t=0$.

In order for our approximation to be self-consistent we should have
$\alpha (l) \ll 1$ for all values of $l$. While the solution
(\ref{alpha-small-angl-sol}) is valid only for $(l_{0}-l) \ll 1$ we assume
it does not differ much from an exact solution even for $l$ close to zero.
We cannot prove this assumption analytically, though a numerical
simulation of Eq. (\ref{alpha2-small-angl}) supports it. Thus, from the
condition of self-consistency for $l=0$ we should have
$\alpha_{0}\sqrt{l_{0}^2/I_{0}+1} \ll 1$. This means that, in the case of
$l_{0}=1/2$, the angle $\alpha$ is always small if $\alpha_{0} \ll
1/\sqrt{l_{0}^2/I_{0}+1}= 1/2$ (we note that this limit is close to $\pi/6
= 30^{\circ}$). For these small initial angles the center of mass of the
pen will pass across the contact point with the finger. In contrast, for
larger initial angles the pen will fall on the side where its center of
mass was located initially. So, the critical angle is in between and it is
larger than zero.


\section{Summary}

We have proposed a simple experiment, which needs a pen and a finger all
in all, and we have formulated it into a mathematical problem. The key
question of the problem is the value of the critical angle which separates
two different types of trajectories of the falling pen. Numerical methods
allow to solve this problem and to find critical angles for any initial
center of mass position $l_{0}$. Theoretical analysis of the equations
provides us the critical angle value in the limit of small $l_{0}$ and
allows us to say that for $l_{0}$ close to $1/2$ the critical angle should
have a finite non-zero value.

We would like to refer the reader to several problems which are also about
the dynamics of the falling pen, however, in a totally different setup.
\cite{Ch1, Ch2, Turner} We have encountered them while studying our
problem. They are quite nice!

\section{Acknowledgements}

We would like to thank Maxim Kirillov for discussions of a numerical
solution, Florian Marquardt for remarks on the manuscript and many other
people for their interest to the problem.

\appendix
\section{An alternative derivation of equations of motion}
\label{alt-way}

We keep the equation which describes the evolution of the angular momentum
in the center of mass frame of reference:
\begin{equation}\label{ang-mom-center-append}
    \frac{d}{dt}\left[I_{0} \omega(t)\right]= N(t)l(t).
\end{equation}
It permits to express the normal force through $\alpha$ and $l$ as
$N=I_{0}\alpha''/l$.

We write then two equations for the evolution of the pen momentum in
projection on $x$ and $y$ axes:
\begin{align}
    m x'' &= N \cos\alpha - \mu N \sin\alpha, \label{x2deriv} \\
    m y'' &= m g - N \sin\alpha - \mu N \cos\alpha. \label{y2deriv}
\end{align}

Next, we introduce polar coordinates according to: $x = - l \sin\alpha$
and $y = - l \cos\alpha$. We note the expressions for the second order
derivatives in these coordinates:
\begin{align}
    -x'' &= l'' \sin\alpha + 2l'\alpha'\cos\alpha
    + l\left(\alpha''\cos\alpha - (\alpha')^2\sin\alpha\right), \nn\\
    -y'' &= l'' \cos\alpha - 2l'\alpha'\sin\alpha
    - l\left(\alpha''\sin\alpha + (\alpha')^2\cos\alpha\right) \nn.
\end{align}


We make variables dimensionless as we did in Sec. \ref{equations} and
finally we obtain Eqs. (\ref{alpha2}) and (\ref{l2}) from Eqs.
(\ref{x2deriv}) and (\ref{y2deriv}) rewritten in polar coordinates.




\end{document}